\documentclass[twocolumn]{aastex631}

\received{April 23, 2025}
\revised{October 28, 2025}
\accepted{November 16, 2025}

\submitjournal{ApJ}

\begin{document}

\title{Stellar Populations with MaNGA: Iron Kink and Nitrogen Fuzz}

\author[0000-0003-1388-5525]{Guy Worthey}
\affiliation{Washington State University \\
Department of Physics and Astronomy, Webster Hall\\
100 Dairy Road Room 1245, Pullman\\
WA 99164}

\author[0000-0002-3077-4037]{Tathagata Pal}
\affiliation{Astrophysics Science Division \\
NASA Goddard Space Flight Center\\
8800 Greenbelt Rd.\\
MD 20771}



\begin{abstract}

Recent analysis of 2968 MaNGA early type galaxies has yielded two notable trends with velocity dispersion ($\sigma$) not previously discussed in the literature. First, Fe abundance rises with $\sigma$, but only until $\sigma\approx100$ km s$^{-1}$, after which it falls. This kink is reproduced by TNG100 simulations, implying that hierarchical merger processes might explain it. Second, astrophysical scatter in N is high for galaxies with $\sigma < 100$ km s$^{-1}$. Due to the restricted list of nucleosynthetic sources for N, it is likely that asymptotic giant branch stars provide most of this N. A varied star formation history (compared to that of massive galaxies) along with variable retention and recycling of N-enriched gas might explain the fuzz of N abundance in low-$\sigma$ galaxies. Because a timescale argument seems necessary to explain the nitrogen fuzz, and an initial mass function argument is ruled out, similar timescale arguments for the [Mg/Fe] trend as a function of velocity dispersion are supported.

\end{abstract}



\section{Introduction} \label{sec:intro}

Absorption line strengths measured in the integrated starlight of galaxies may be used to infer some limited information about their chemical evolution and star formation histories \citep{2000AJ....119.1645T,2012MNRAS.421.1908J,2014ApJ...780...33C,2014ApJ...783...20W}. Recently, the inclusion of isochrones that respond to individual element abundance changes \citep{2022MNRAS.511.3198W} has shifted the measurement of nonsolar abundance ratios from integrated light, which had heretofore been heuristic, to a proper scale relative to the solar mixture. That is, derived quantities such as [Mg/Fe] or [Fe/H] can now reasonably be expected to be free of zero point shifts and spurious slope changes wrought by hybrid stellar population models whose spectra shift with elemental mixture but whose underlying isochrones remain static. In this happier situation, it behooves us to reexamine some of the excellent spectra that exist for galaxies.

\section{Data, Analysis, and Results} \label{data}

\citet{PalWorthey25} analyzed 2968 MaNGA low redshift galaxies from DR17 \citep{2022ApJS..259...35A} with S\'ersic index $>2$, so that they are, structurally at least, early-type galaxies (ETGs). The spectra have a resolution of $R\sim2000$ covering a $3000 < \lambda < 10000$ \AA . Any galaxy with significant H$\alpha$ emission was dropped from the sample in order to make mean age estimation more accurate. Elliptical annuli were superimposed to extract radially dependent spectra. 

An array of absorption feature indices measured from the galaxy spectra were compared to stellar populations models. The model grid is that of \citet{2022MNRAS.511.3198W} using an isochrone set assembled from corrected and updated versions of the \citet{2008A&A...482..883M} grid, a \citet{2001MNRAS.322..231K} initial mass function, and a library of spectral indices averaged over four large libraries \citep{2004ApJS..152..251V,2006MNRAS.371..703S,2014A&A...561A..36W,2009ApJS..185..289R} with departures caused by changes to chemical composition calculated via synthetic spectra. Single Stellar Populations (SSPs) were subsequently integrated over an analytic abundance distribution (ADF) function of `normal' width \citep{2014MNRAS.445.1538T}. The resulting composite stellar populations (CSPs) were characterized by single-burst age, the [Fe/H] value of the \textit{peak} of the ADF, and elemental abundances [C/Fe], [N/Fe], [Na/Fe], and [Mg/Fe]. The philosophy for this MaNGA reduction pass was to derive high-accuracy abundances from easy-to-measure features. For example, oxygen abundance is strongly degenerate with derived age \citep{2022MNRAS.511.3198W} but oxygen's spectral impact is not secure. Oxygen and [Si/Fe] were assumed to vary in lockstep with [Mg/Fe]. A model-inversion program called {\sc compfit} \citep{2014ApJ...783...20W,2022MNRAS.511.3198W,2023MNRAS.518.4106W} searched for best-fit ages and overall abundance for a trial chemical mixture. The element mixture was altered via linear fits to sensitive features, then the process was iterated. Errors in astrophysical parameters were estimated by altering the input galaxy indices within error envelopes and recomputing. Fifty such Monte Carlo realizations were computed for each annulus in each galaxy.

The CSPs we use are ``one known bias better'' than SSPs along the metallicity dimension, but for this reduction run we still output a single age for systems which, presumably, have much more complex star formation histories. The single age we adopt is approximately equivalent to a photometrically-weighted mean age, since young populations are brighter. Na\"{\i}vely, a $U$- or $B$-weighted mean age might be equivalent, since our age-sensitive features (Balmer features) are near 4000\AA , but tests indicate that a passband slightly blueward of $U$ is the best at reproducing the mean age produced by Balmer indices.

A more complete procedural description and the full table of results is given in \citet{PalWorthey25}. From that paper, however, we would like to highlight two observations that might advance our understanding of galaxy chemodynamical evolution. 

\subsection{Nitrogen Fuzz}

\begin{figure}[h]
\plotone{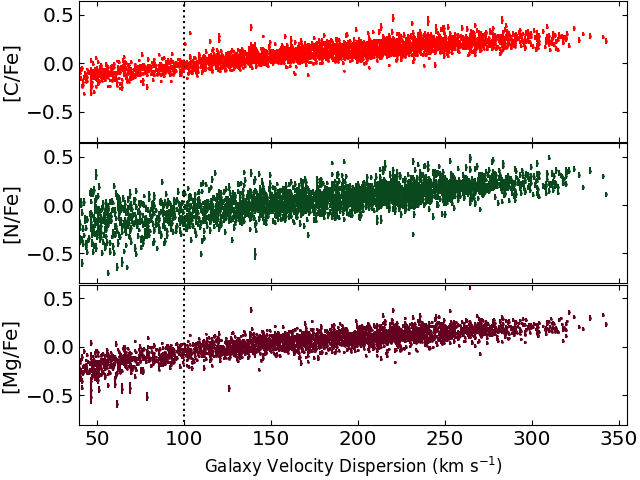}
\caption{Light element abundances relative to Fe as a function of velocity dispersion. Errorbars are Monte Carlo estimates based on random observational noise within a 0.7 Gaussian $\sigma$ envelope. Galaxies with $\sigma < 100$ km s$^{-1}$ show pronounced scatter in [N/Fe] that is not seen in [C/Fe] or [Mg/Fe].
\label{fig:light}}
\end{figure}

\begin{figure}[h]
\plotone{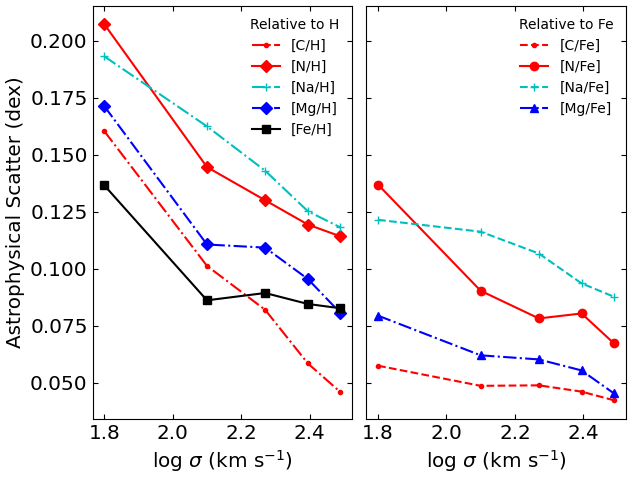}
\caption{Astrophysical scatter in logarithmic abundances binned by galaxy velocity dispersion. The smallest galaxies have the largest scatter in all quantities, but nitrogen's scatter has the greatest magnitude.
\label{fig:astroscat}}
\end{figure}

For galaxies with $\sigma < 100$ km s$^{-1}$, the scatter in [N/Fe] is large compared to that in [C/Fe] or [Mg/Fe], as illustrated in Fig. \ref{fig:light}. The abundances plotted are from the 2nd annulus centered at $R = 0.35 R_e$. Fig. \ref{fig:astroscat} shows that the scatter is not observational in origin. In Fig.\ref{fig:astroscat}, galaxies are binned by velocity dispersion, then the median observational errors are subtracted in quadrature from the total scatter, leaving an estimate of astrophysical scatter. We chose annulus 2 to be close to a radius characteristic of the galaxies' velocity dispersion, but the result is the same, no matter which annulus is examined.

Low-$\sigma$ galaxies tend to have flat abundance gradients, even in Fe and Na, which two elements show strong gradients in high-$\sigma$ galaxies \citep{PalWorthey25}. The large [N/Fe] variation from galaxy to galaxy applies therefore to the bulk of the mass in these small galaxies. Reading from Fig. \ref{fig:astroscat}, variations in the histories of these galaxies engender a factor of $\pm40$\% variation in [N/Fe] (one sigma) whereas [C/Fe] or [Mg/Fe] vary by less than 20\%.

We searched for parameter correlations that might shed light on [N/Fe]'s behavior, including correlations with age residuals from the age-$\sigma$ relation. Restricted to the 501 galaxies with $\sigma < 100$ km s$^{-1}$, [N/Fe] seems to operate independently of other parameters. For example, the coefficients of determination $R^2$ for [N/Fe] or [N/Mg] versus age or abundance are all less than 0.05, that is, largely uncorrelated, whereas [Mg/Fe] against log age has $R^2 = 0.41$ and [Fe/H] against log age has $R^2 = 0.22$. All [X/H] combinations correlate well with each other.

The lack of an age correlation is not very meaningful, since a mean age is far from what we really wish to measure, which is the timescale over which gas is retained and recycled into stars. Also, age has the largest scatter of any of the astrophysical parameters we attempt to measure. Only for low-$\sigma$ galaxies can the age scatter be shown with security to be astrophysical rather than observational.

\subsection{Iron Kink}

\begin{figure}[h]
\plotone{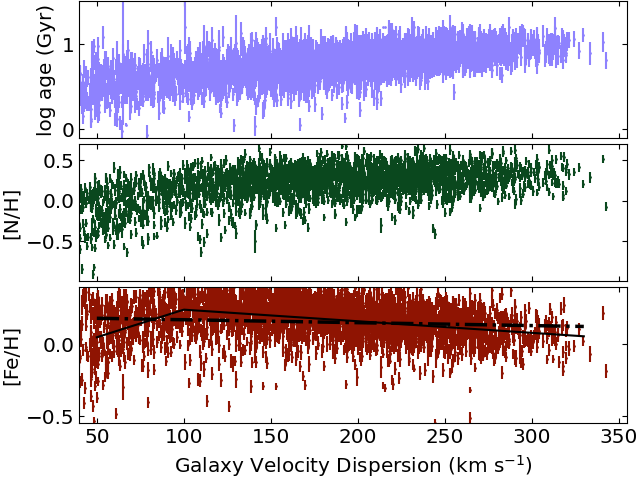}
\caption{Log age and abundances for N and Fe as a function of $\sigma$ with errorbars as in Fig. \ref{fig:light}. One-error regressions for a line and a broken line are overplotted in the Fe panel. The top of the Fe panel is set to [Fe/H] = 0.4, the limit imposed by the model grid.
\label{fig:ageiron}}
\end{figure}

Astrophysical scatter is highest in \textit{all} derived quantities for smallest $\sigma$ \citep[Fig. \ref{fig:astroscat}]{PalWorthey25}, including mean age, illustrated in Fig. \ref{fig:ageiron}. The mean age for small galaxies tends to be about half the age of the universe, perhaps implying a more extended (or less quenched) star-formation history compared to giant ETGs. 

Also shown in Fig. \ref{fig:ageiron} is the behavior of [Fe/H], which rises with $\sigma$ until $\sigma = 100$ km s$^{-1}$ and then falls. The most Fe-rich galaxies have $100 < \sigma < 120$ km s$^{-1}$, while for larger galaxies it is the light metals that drive increasing heavy element abundance. So, giant ellipticals are the most metal-rich, but also the most Fe-poor of ETGs. An F-test on the residuals from the unbroken and broken line fits indicates that the broken line fit is highly significant (insignificance probability $10^{-14}$ for the annulus illustrated in Fig. \ref{fig:ageiron}).


Finally, Fig. \ref{fig:ageiron} shows [N/H], which does not fall at high $\sigma$ like [Fe/H]. Therefore, if the cause of Fe's fall is that light metals from Type II supernovae are diluting the most massive of galaxies, then N has a compensatory source of enrichment not shared by Fe. This is likely to be AGB stars, as discussed next.

\section{Discussion}

The iron kink and the nitrogen fuzz are surely clues to the details of galaxy chemodynamical evolution. As presented in some detail by \citet{PalWorthey25}, the TNG100 simulations \citep{2018MNRAS.480.5113M, 2018MNRAS.477.1206N, 2018MNRAS.475..624N, 2018MNRAS.475..648P, 2018MNRAS.475..676S} reproduce the Fe kink at almost exactly the same point ($\sigma\approx 100$ km s$^{-1}$). This agreement is convincing because neither the TNG authors nor we sought to replicate such a thing. The qualitative agreement therefore lends strong support to the hierarchical assembly picture of galaxy evolution in that it \textit{can}, demonstrably, reproduce this relatively subtle chemical signature.

The nitrogen fuzz, on the other hand, is novel. The TNG100 yield assumptions are generally too high for the observations we present, although both for them and us the $\sigma = 100$ km s$^{-1}$ point is an inflection point in slope for all quantities except mean age, especially apparent when plotted versus log~$\sigma$.

\begin{figure}[h]
\plotone{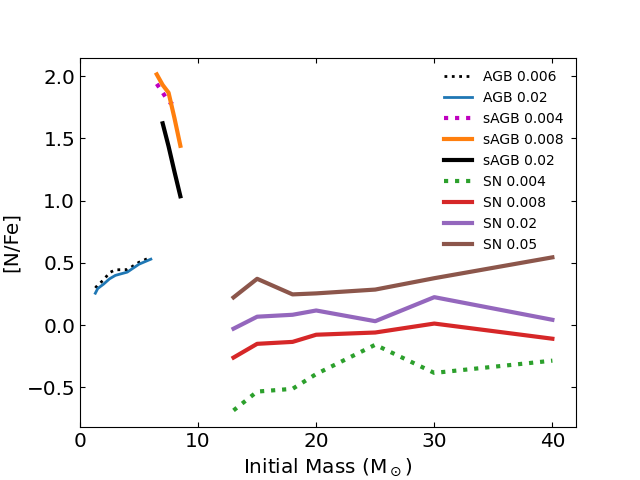}
\caption{Nitrogen yield relative to iron as a function of initial stellar mass for asymptotic giant branch (AGB) stars \citep{2011ApJS..197...17C} at low masses, super-AGB stars \citep{2014MNRAS.437..195D} at intermediate masses, and core-collapse supernovae \citep{2013ARA&A..51..457N} at high masses. Colors indicate initial progenitor heavy element mass fractions ($Z$) as marked.
\label{fig:yields}}
\end{figure}

According to \citet{2013ARA&A..51..457N}, neither hypernovae, pair-production supernovae, Type Ia supernovae, nor neutron star collisions yield significant amounts of N. Nitrogen therefore has two main nucleosynthetic sources, asymptotic giant branch (AGB) stars and core-collapse supernovae (Fig. \ref{fig:yields}). 

The supernova yield predictions in Fig. \ref{fig:yields} from \citet{2013ARA&A..51..457N} show a weak dependence on the progenitor mass but a strong dependence on the progenitor metallicity. While noting that supernova yields are notoriously difficult to predict, it seems from the figure that only chemically evolved gas can produce `instant' N-rich ejecta. The AGB stars, on the other hand, eject much N, but over timespans of hundreds of millions of years after formation. Therefore, galaxies that quench effectively should not be able to take advantage of the AGB star contributions. Galaxies that have prolonged star formation histories and can retain their gas, however, could be enriched in N, albeit at the whims of gas geometry and the timing of individual bursts of star formation.

Low-$\sigma$ galaxies are the ones with the youngest mean ages, and hence the most evidence for protracted star formation histories. These galaxies are also too dynamically cold to generate coronal, X-ray-emitting gas, and therefore lack that mechanism for star formation quenching. Low-$\sigma$ galaxies are also the ones that exhibit ``strong N individualism,'' or, in other words, [N/Fe] fuzz.

Iron comes in large part from Type Ia supernovae, which require white dwarfs to trigger. Therefore, they come with an inherent time delay, like AGB star enrichment. If the authors could arrange the universe to their liking, the Type II enrichment would be nearly instant, the Type Ia would follow after some $10^8$ yr, and then AGB enrichment would linger the longest. In that case, Mg (Type II), Fe (Type Ia), and N (AGB star) abundances could indicate how long gas lingers in a galaxy before being recycled into stars. Alas, famously, the Type Ia enrichment timescale is not known very well.

\begin{figure}[h]
\plotone{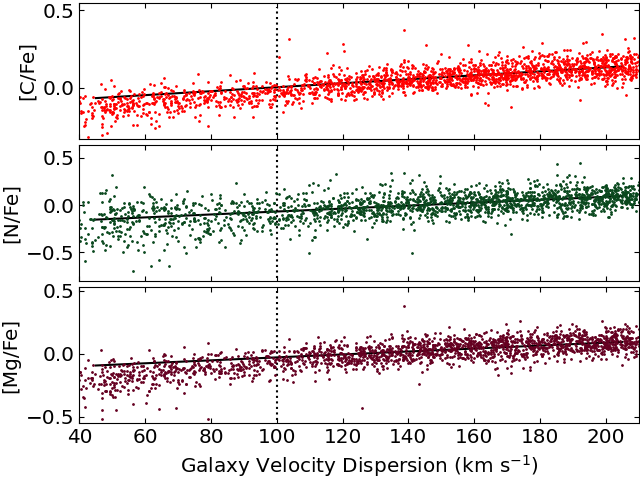}
\caption{[C/Fe], [N/Fe], and [Mg/Fe] versus velocity dispersion. This version of Fig. \ref{fig:light} suppresses error bars and zooms to a narrower velocity dispersion range. A line fit over data in the range 110 km s$^{-1} < \sigma > 340$ km s$^{-1}$ is superimposed (solid line). Note that for galaxies $< 100$ km s$^{-1}$ (to the left of the dotted line), [C/Fe] and [Mg/Fe] lie low, as if these galaxies were enhanced in Fe, but [N/Fe] follows the trend set by high-$\sigma$ galaxies. 
\label{fig:departure}}
\end{figure}

Fig. \ref{fig:departure} points out possible support for such a measurement by fitting lines to [X/Fe] versus $\sigma$, but only large ones with  110 km s$^{-1} < \sigma < 340$ km s$^{-1}$. The extrapolations at low $\sigma$ miss [C/Fe] and [Mg/Fe], as predicted if extended star formation histories favor Fe over Type II supernova products. But N should also arise plentifully in extended star formation scenarios, and we see it follow the extrapolation of the Fe-N trend. N is being incorporated into stars at about the same rates as Fe for these low-$\sigma$ galaxies. 

This might finger timescales as the driving mechanism, but it might not. \cite{1992ApJ...398...69W} also list initial mass function (IMF) as a possible cause. Star formation in massive galaxies should be more dynamically turbulent, and this may favor the formation of massive stars, say 50 M$_\odot$ relative to 10 M$_\odot$, producing more Mg (and C). The behavior in Fig. \ref{fig:departure} might just as easily be caused by an IMF-caused dearth of Type II supernova products. The scatter in [N/Fe], however, is real, and there must be a cause, and it cannot be IMF since white dwarfs and AGB stars have similar initial masses. 

The nitrogen fuzz therefore gives us indirect support for timescale arguments in chemical enrichment scenarios. While these timescale arguments have long been favorites \citep{2002Ap&SS.281..371T,2012MNRAS.421.1908J,2019MNRAS.483.3420P}, it is nice to have a reason to prefer them that is grounded in data.

Future work on the topics addressed in this paper should include attempts to go beyond ``mean age'' for each galaxy. A more direct connection between N buildup and protracted star formation may then be established. Too, measurement of oxygen abundances from the existing MaNGA spectra would inform age, via the oxygen-age degeneracy. Inclusion of blue straggler stars in the stellar population models should occur before age zero points are taken seriously.

Because high redshift dwarf galaxies are difficult to observe, the Fe kink might be best addressed using numerical galaxy formation codes to discover which assumptions and initial conditions affect its successful prediction.

\begin{acknowledgments}
This research was supported by the NASA Postdoctoral Program at Goddard Space Flight Center, administered by Oak Ridge Associated Universities under contract with NASA.
\end{acknowledgments}

%





\bibliography{kinkfuzz}{}
\bibliographystyle{aasjournal}

\end{document}